\documentstyle[11pt,newpasp,twoside,epsf]{article}
\def\edcomment#1{\iffalse\marginpar{\raggedright\sl#1\/}\else\relax\fi}
\reversemarginpar

\begin{document}
\title{Evolution of Brown Dwarf Atmospheres: Investigating Physical Parameters from Near-IR Spectra.}
 \author{Nadya Gorlova\altaffilmark{1}, Michael R. Meyer, Jim Liebert, George H. Rieke }
 \affil{Department of Astronomy, The University of Arizona, Tucson, AZ, USA 85721}
\altaffiltext{1}{Email: ngorlova@as.arizona.edu}

\begin{abstract}
We obtained near--infrared spectra of a sample of 
very low mass objects as a function of age in order 
to investigate the temperature and surface gravity 
sensitivity of several features in the J-- and K--bands.
\end{abstract}

\section{Introduction}

In order to study the initial mass function in young open clusters at the low mass end,
one needs to have a fast and reliable way of separating pre-MS members
from foreground dwarfs and background giants. These objects occupy
same place on the color-magnitude diagram, but have different surface gravities.
Using this fact, we decided to explore the  gravity-sensitive features in the near-infrared 
spectra of the objects of spectral types M--L, corresponding to the range of masses 
near sub-stellar boundary.

\section{Observations}

Using FSpec IR spectrograph on the 6.5m MMT telescope (Mt. Hopkins, Arizona),
we obtained  J \& K spectra of $\sim$ 30 brown dwarf candidates, SpT M4--L5, drawn from the field
and open clusters and associations ($\rho$ Oph, Tau, IC 348, Upper Sco, Pleiades) 
to sample a range of gravities $logg$ $\sim$ 3 to 5 dex (Fig.1), as well as 3 late M giants ($logg$$\sim$0).
The resolution of our smoothed spectra is $\sim$700 and an average S/N in flux is 40--80
(depending on the feature).

\section{Results}

We calculated gravities using masses from the theoretical
$T_{eff}$--$L$ diagram, assuming dwarf colors and having distances to most
of our objects. The correlation of some features measured from our spectra with gravity is shown in Fig.1.

We confirm strong temperature dependence of H$_2$O and CO bands (eg. Reid et al. 2001)
and infer that gravity dependence of these features is insignificant
compared to that of FeH, KI and NaI lines. While the weakness of the latter features 
in cluster compared to field objects was known before (eg. Martin et al. 1996, 
Luhman et al. 1998, Bejar et al. 1999, Lucas et al. 2001), no systematic investigation of this effect has been
carried out so far, especially in the J band. IR is the only option to work with cool, low mass
and embedded objects, and in the J band the atomic features are especially prominent.

We also compared our spectra to recent atmospheric calculations by Allard et al. 2001
with the conclusion that while they predict correct trends, the systematic offsets
between observations and theory remain (Fig.1).
  
Using our results we managed to confirm low surface gravity 
of two brown dwarfs from the TW Hya and Taurus associations, whose membership 
was originally proposed based on their optical spectra (2MASS1139-3159 and CFHT-BD-Tau4).

\begin{figure}[b]
\plotfiddle{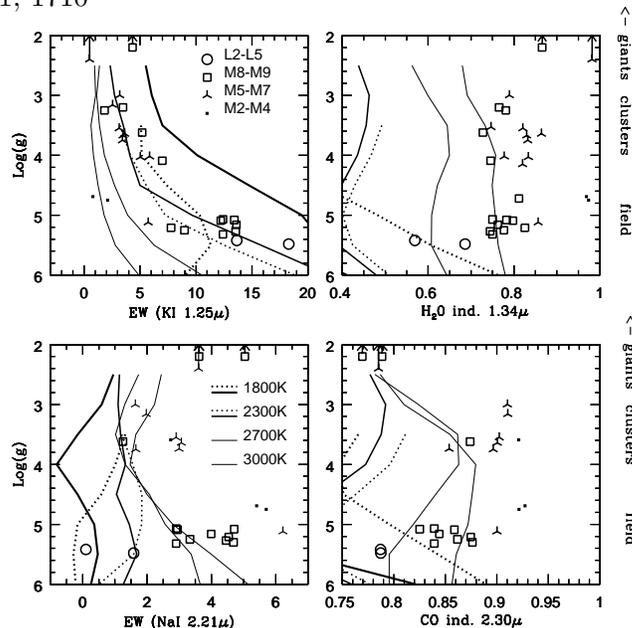}{0pt}{0}{43}{43}{-140}{-83}
\caption{The trends of the strength of some absorption features in the near-IR with gravity and temperature,
as measured from our spectra (symbols) and PHOENIX model spectra (dots-- Dusty2000 and solid-- Cond2000). Giants 
are indicated by arrows as their surface gravities are off scale
on this plot ($logg$$\sim$0).} 
\end{figure}

\end{document}